\documentclass[twocolumn]{article}
\usepackage{authblk}
\usepackage{graphicx}
\usepackage{hyperref}
\usepackage[normalem]{ulem}
\usepackage{color}
\usepackage[usenames,dvipsnames,svgnames,table]{xcolor}

\title{Single-shot X-ray Absorption Spectroscopy at X-ray Free Electron Lasers}

\author[1]{ M. Harmand }
\author[2]{ M. Cammarata}
\author[3]{ M. Chollet }
\author[1,4]{ A.G Krygier}
\author[3,5]{ H.T. Lemke }
\author[3]{ D.Zhu}

\affil[1]{ IMPMC, Sorbonne Universit\'e, UMR CNRS 7590, MNHN, 75005 Paris, France }
\affil[2]{Institut de Physique de Rennes, UMR UR1-CNRS 6251, Universit\'e de Rennes 1, F35042, Rennes, France }
\affil[3]{ LCLS, SLAC National Accelerator Laboratory, Menlo Park, CA 94025, USA }
\affil[4]{ Lawrence Livermore National Laboratory, 7000 East Ave, Livermore, CA 94550, USA }
\affil[5]{ SwissFEL, Paul Scherrer Institut, Villigen 5232, Switzerland }

\date{\today}

\begin{document}

\twocolumn[
\begin{@twocolumnfalse}
\maketitle

\begin{abstract} 

X-ray Absorption Spectroscopy (XAS) is a widely used X-ray diagnostic method. While synchrotrons have large communities of XAS users, its use on X-Ray Free Electron Lasers (XFEL) facilities has been rather limited. At a first glance, the relatively narrow bandwidth and the highly fluctuating spectral structure of XFEL sources seem to prevent high-quality XAS measurements without accumulating over many shots. Here, we demonstrate for the first time the collection of single-shot XAS spectra on an XFEL, with error bars of only a few percent, over tens of eV. We show how this technique can be extended over wider spectral ranges towards Extended X-ray Absorption Fine Structure (EXAFS) measurements, by concatenating a few tens of single-shot measurements. Such results open indisputable perspectives for future femtosecond time resolved XAS studies, especially for transient processes that can be initiated at low repetition rate.\\

\end{abstract}
\end{@twocolumnfalse}
]

\section{Introduction}

X-ray Absorption Spectroscopy (XAS) is an extremely powerful x-ray diagnostic for simultaneously probing the electronic structure and the local atomic arrangement. XAS is a widely used technique and its community has been continuously growing since the first quantitative work exploiting x-ray synchrotron radiation, almost 50 years ago. XAS is now used for a wide range of scientific areas such as fundamental physics, material science, plasma physics, biology, geoscience, chemistry and cultural heritage \cite{Abe2008}. More recently, the development of time-resolved XAS, often coupled to pump-probe techniques, has allowed tracking of the ultrafast interplay between electron and atomic motion or, opening new perspectives for understanding complex processes \cite{Chergui2016, Dorchies2016}. 

In this context, the unique properties of X-ray Free Electron Laser (XFEL) sources now makes it possible to apply this now traditional X-ray synchrotron technique to a wider range of systems that were not previously accessible. There are already several XFEL facilities operating around the world that are capable of routinely producing extremely short pulse lengths on the order of 1 - 10 femtoseconds (fs) and below \cite{Kang2020}, tunable photon energy and extremely high intensity (up to 10$^{20}$ W/cm$^{2}$). Such properties are of great interest for a wide range of applications and part of this success is due to the ability of researchers to adapt advanced and well developed synchrotron x-ray techniques that particularly benefit from higher photon brilliance at the fs timescales. For example, a concerted effort of many research groups has paved the way to the development of serial femtosecond crystallography based on fs x-ray diffraction snapshot of thousand of micron size crystals \cite{Boutet2012}. Ultrafast single-shot x-ray diffraction measurements of laser compressed samples are also now commonly performed at XFEL facilities allowing pump - probe measurements of transient phases and  phase transition mechanisms \cite{Gleason2015, McBride2019}. The fs pulse duration of the XFEL allows probing of the atomic arrangement before any atomic motions induced by the heating of the XFEL probe \cite{Inoue2016}. In addition, the high brightness of the XFEL pulse often enables (quasi-)single-shot measurements, ensuring better statistics, higher reliability of the data and time for scanning wide conditions during an experiment. 

While XAS is one of the most powerful synchrotron x-ray diagnostics for studying the interplay between the electronic structure and the local atomic arrangement, its implementation at XFEL facilities has so far been limited. The main barrier to performing XAS with XFEL sources arises as a natural side effect of how the pulses are generated. The self-amplified spontaneous emission (SASE) process, currently used in all hard X-ray facilities, results in high XFEL brightness but yields poor source longitudinal coherence and highly fluctuating spectral structure making XAS measurement particularly difficult. The favored approaches for overcoming this issue consists of using a monochromator or the seeded operation mode \cite{Amann2012, Kroll2016} to select and scan the photon energies across edges and resonances \cite{Lemke2013}. This comes at an additional cost of even higher fluctuations and limited photon flux due to the 95$\%$ loss the monochromator. As a result, scanning XAS has so far not been possible without accumulating over tens of thousands of repeatable shots. In the case of pump - probe measurements, this becomes even more critical as it also implies accumulating pump - probe pulse pairs at a given delay, then assuming high stability and repeatability of both the sample excitation and probed conditions during the whole data acquisition. Consequently, this imposes strong experimental restrictions as well the variety of conditions that can be probed during a limited time. Moreover, for many experiments, high repetition rate accumulations are not possible. This is especially true for the irreversible processes, for example occurring in studies of materials in extreme conditions where the shot-to-shot fluctuations or transient effects can be the subject of the study and the required data accumulation or x-ray energy scans are impossible or highly limited. 
To overcome these limits, alternate dispersive methods allow collecting an entire spectral range at once, using dispersive optics and position sensitive detectors. Such approach has been initially developed at synchrotrons \cite{Pascarelli2016, Torchio2016, Mathon2015} and more recently at XFELs  \cite{Gaudin2014, Harmand2015}. In these cases, it has consisted of using a dispersive XAS setup coupled with a specific data treatment to enable accumulating over ten's of shots to reach several percent signal to noise ratio. Unfortunately, this technique developed so far still requires assuming sufficiently identical experimental conditions, which in the best case is not ideal and can also place undesirable constraints on the experimental conditions and configurations, even at the cost of losing a significant amount of data \cite{Harmand2015}. In this work, we demonstrate a new approach to performing ultrafast single-shot X-ray Absorption Near Edge Spectroscopy (XANES) measurements at XFEL facilities. We also show the capability to perform Extended X-ray Fine Absorption Spectroscopy (EXAFS) by accumulating tens of shots.
 
 \begin{figure*}[t]
 \centering
 \includegraphics[width=6in]{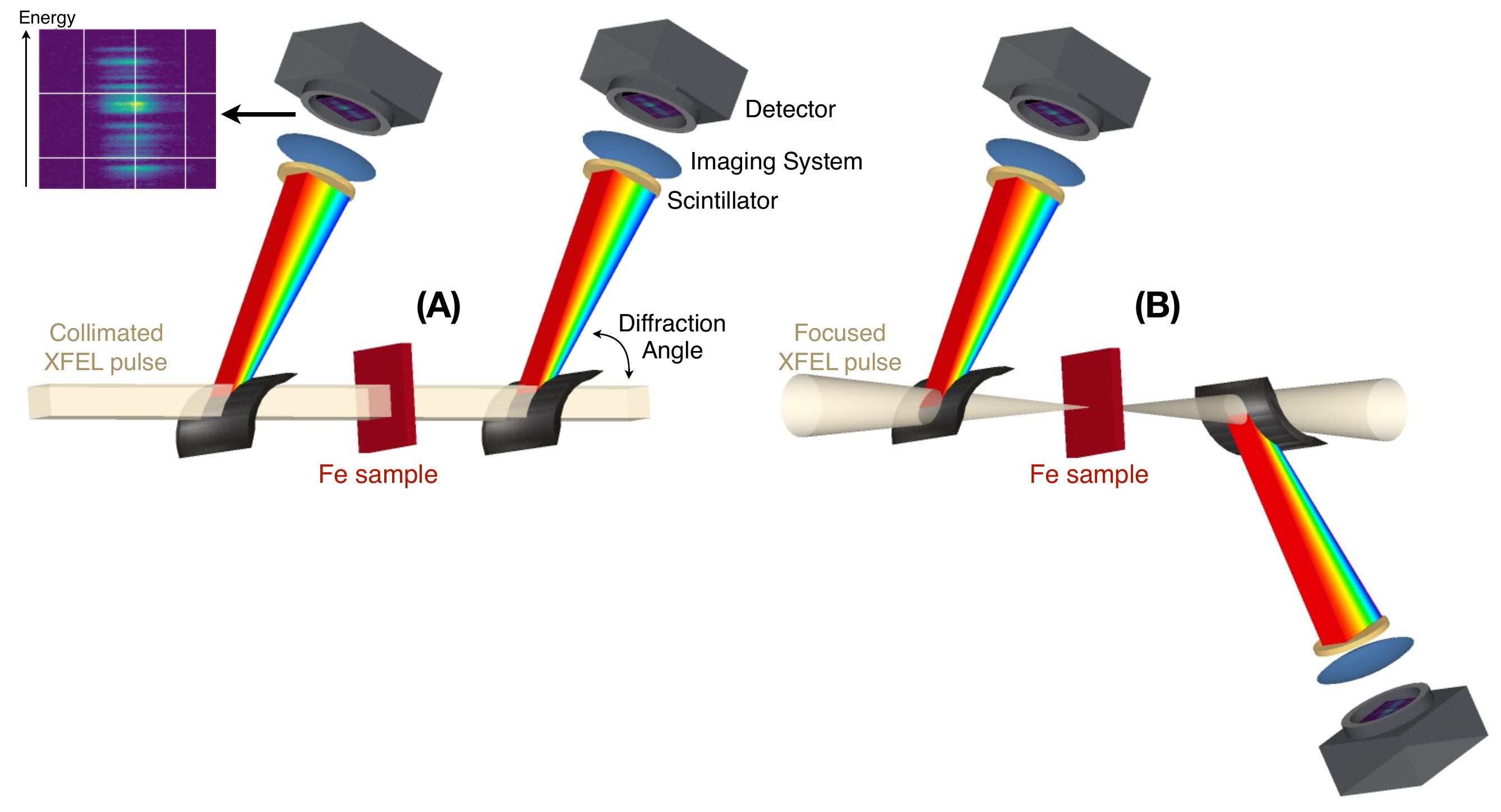}
 \caption{Energy dispersive experimental setup at LCLS with the unfocused collimated (A) and the focused (B) XFEL beam. Two identical spectrometers allow simultaneous measurement of the incident and transmitted spectra. }
 \label{Setup}
\end{figure*}

\section{Results}

\begin{figure}[!h]
 \centering
 \includegraphics[width=3in]{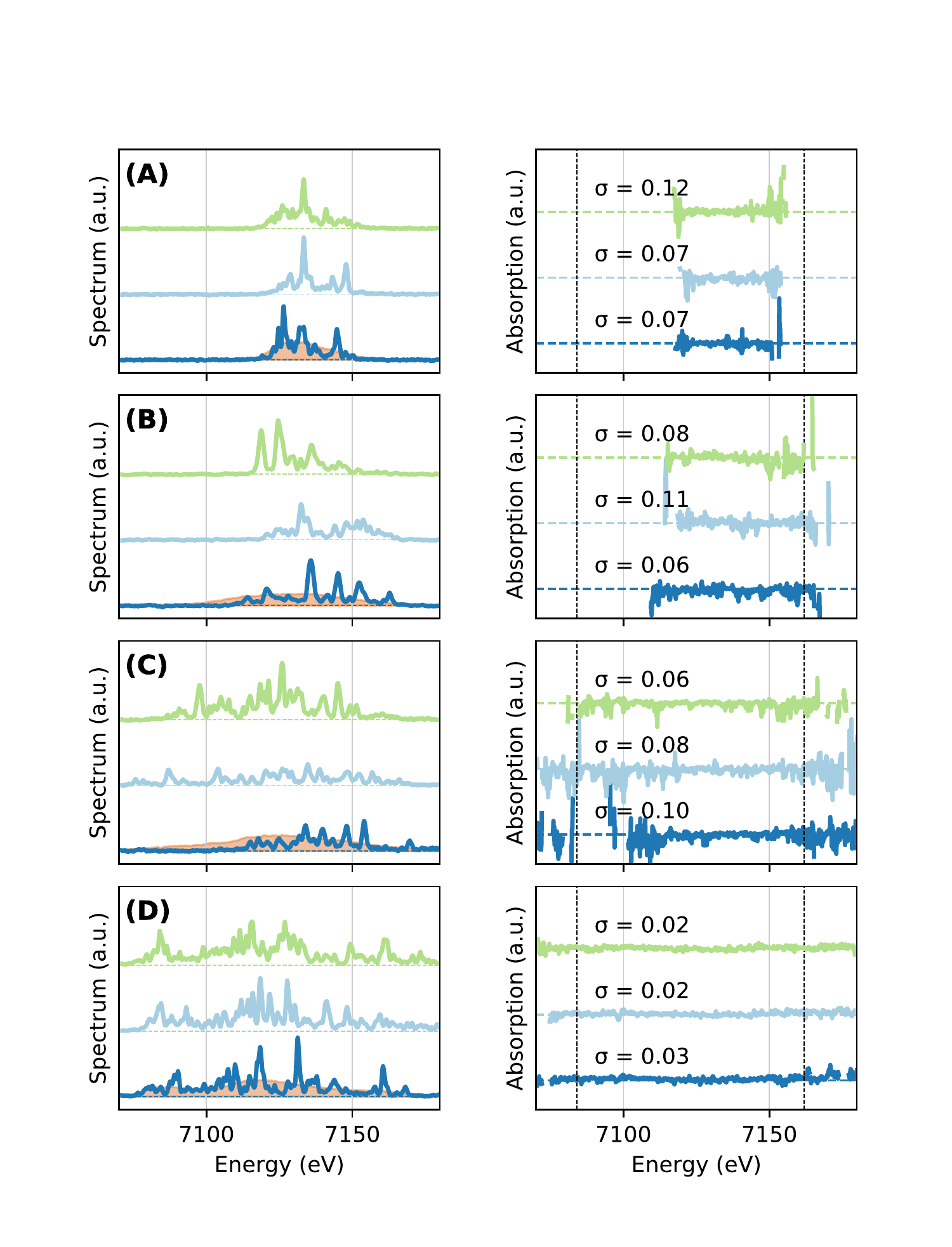}
 \caption{Left - Three consecutive single-shot XFEL spectra are displayed in dark blue, pale blue and green. They have been obtained for various tuning of the LCLS machine \cite{Welch2011} : low charge mode \cite{Ding2015} (A), low charge mode with large bandwidth (B) nominal mode full charge (C), overcompressed mode \cite{Ding2016} (D). The 1000-shot average spectra is shown in orange on the left panel. Right - Corresponding single-shot absorption spectra with their calculated standard deviation $\sigma$, showing the fluctuations around the average absorption value. This value is calculated for the spectral range shown by vertical dashed lines. The extraction of the XANES spectra is described in the method section. }
 \label{FELSpectra}
\end{figure}

The presented experiment was performed at the XPP instrument of the LCLS \cite{Emma2004, Chollet2015}. Dispersive XAS measurements require normalizing a measured transmitted spectra $I_{T}$(E) by the incident spectra $I_{0}$(E) measured simultaneously for every single x-ray pulse. The energy dispersive experimental setup (see figure \ref{Setup}) consists of two identical spectrometers \cite{Harmand2015, Zhu2012} in a reference - transmission geometry. The first spectrometer, placed upstream of the sample, collects the reference source spectrum while the second spectrometer, placed downstream of the sample, collects the transmitted spectrum. This experimental setup, combined with a dedicated spectrometer alignment procedure, an advanced data analysis and an optimization of the XFEL spectral properties allows us to overcome the limitations with the typical shot-to-shot spectral and spatial variations that originates from the stochastic Self-Amplified Spontaneous Emission (SASE) FEL processes.  

First, the quality of the absorption spectra relies critically on the relative alignment and calibration of the two spectrometers. Indeed, the dispersive mechanism of the single-shot spectrometer maps different spatial positions of the beam profile to a different photon energy value (along the dispersion direction, vertical in figure \ref{Setup}(A)). An implicit assumption is that the spectral content of the incident beam is spatially uniform. This however, is not always the case for the SASE FEL pulses. Past observation have shown that spectra sampled from different part of the beam can be significantly different. This correlation between the spectral content and the spatial coordinate of the XFEL pulse profile is called ''spatial chirp''. Here, we chose the vertical scattering geometry in which the magnitude of the spatial chirp is expected to be the lowest due to the horizontal dispersion elements in the accelerator \cite{Zhu2013}. In addition, this effect can be further minimized if for each specific photon energy, the two spectrometers sample the spectral intensity from the neighboring spatial region of the beam profile. This is accomplished by scanning the angle of the downstream spectrometer and locating an intensity drop in the second spectra caused by extinction from the crystal of the incident upstream spectrometer (see the Supplementary Material figure 1). This alignment procedure should be performed for each beamline configurations and FEL operation modes.

Once the two identical spectrometer alignment is optimized, the 2D images extracted from each spectrometer still would not match perfectly as systematic 2D mismatchs need to be corrected for. We proceed by first defining a 2D image transformation function $\mathcal{F}_{2D}$ to match the downstream and upstream spectrographs. This function  $\mathcal{F}_{2D}$ is derived from the average of few tens of pulses acquired with no sample and is then applied to the absorption spectra with the sample inserted. $\mathcal{F}_{2D}$ consists of 2D affine transformations such as rotation, translation and shear. In an iterative process, the difference between the two spectrometer images is minimized by adjusting the transformation parameters of $\mathcal{F}_{2D}$ (see Method and Supplementary material). The quality of this iterative procedure is estimated by calculating the absorbance $A$, that should equal 0 at all energies in absence of a sample. $A = -log(T) = \frac{(I_{T})(E)}{I_{0}(E)}$, with $I_{T}(E)$ and $I_{0}(E)$ being the integration of the 2D images, perpendicular to the dispersive direction. The minimized standard deviation $\sigma$ of the no sample absorbance is used to quantify the noise associated with our measurements.  We also found that it is necessary to record the no sample reference spectrographs shortly before inserting the sample for actual data collection. 

Finally, we have explored various XFEL operation modes to directly modify the XFEL spectral characteristics and optimize the single-shot absorption spectra quality. In figure \ref{FELSpectra}, four different machine operation modes are shown and labeled from (A) to (D). Three single shot spectra are displayed for each XFEL configuration. The absorption is calculated after applying the transformation function $\mathcal{F}_{2D}$, optimized on previous shots.  As explained before, we show results with no sample inserted in the XFEL beam, so that the effect on the standard deviation $\sigma$, that directly relates to the XANES measurements noise, is directly visible and quantified. As we can observe, the XFEL tuning strongly impacts the quality and more precisely the signal to noise ratio of the XANES measurement. 

First, the fluctuating XFEL spectral spikes are clearly visible in both the measured reference and transmission spectra (left panels of figure \ref{FELSpectra}). The highly non-uniform distribution of the intensity is a challenge for normalization as for each shot there will be spectral range with very low intensity. Our previous work performing dispersive XANES measurements using XFEL radiation addresses the problem by averaging multiple shots in order to obtain a more uniform distribution of the reference spectra \cite{Gaudin2014, Harmand2015}. In this experiment however, we overcome this challenge by first identifying and optimizing the XFEL operation mode to increase the spectral bandwidth and more importantly minimize the spectral regions with few photons, i.e. ensuring a minimum level of counting at each energy of the spectra. This XFEL machine optimization has been performed by the accelerator experts of LCLS and consists in optimizing different modules of the Linac that directly affect the properties of the electron bunch spectra and therefore the single-shot X-ray photon spectra. 

In the normal operation mode, the SASE bandwidth is determined largely by the gain length and the spectral full widths are typically 0.2-0.3$\%$, or 15-20~eV at 7~keV \cite{Ding2016}. One way to increase the overall bandwidth of the X-ray pulses is to increase the energy-chirp of the electron bunches \cite{Guetg2016, Guetg2016b}. In practice, the over-compression mode, with electron bunch head-tail switches in the compressor chicane of the accelerator \cite{Ding2016, Turner2014}, produces a reversed energy correlation along the length of the electron bunch (lower energy at the tail side) known as chirp. Linac structure wakefields downstream of the compressor chicane introduce additional energy loss in the tail part of the electron bunch, which further enhances the amplitude of the existing energy chirp. By spreading the electron energy, different slices of the bunch in the undulator can therefore lase independently at different photon energies leading to increased XFEL bandwidth \cite{Welch2011} and to ensure a non-zero X-ray photon distribution over the whole bandwidth. In particular the `over-compressed' mode allows us to extend the spectral range to almost 100 eV, more than 5 times of the typical SASE bandwidth, as shown in Figure \ref{FELSpectra}-D.  Here, the overcompressed mode is systematically showing a standard deviation $\sigma$ below 0.05 and down to 0.02 - 0.03 in most of the cases. This implies $\sim ~ 2-3\%$ noise associated with our single-shot absorption measurements. This mode has been identified as the most suited one for single-shot XANES measurements at LCLS. Other modes such as the low charge mode, with and without enlarging of the bandwidth or the nominal mode full charge (figure \ref{FELSpectra}-A, -B and -C respectively) produce between $\sim$~4 and 10$\%$ noise in average. All data presented below have been obtained with the overcompressed mode. \\

\begin{figure*} [!h]
\centering
 \includegraphics[width=5.5in]{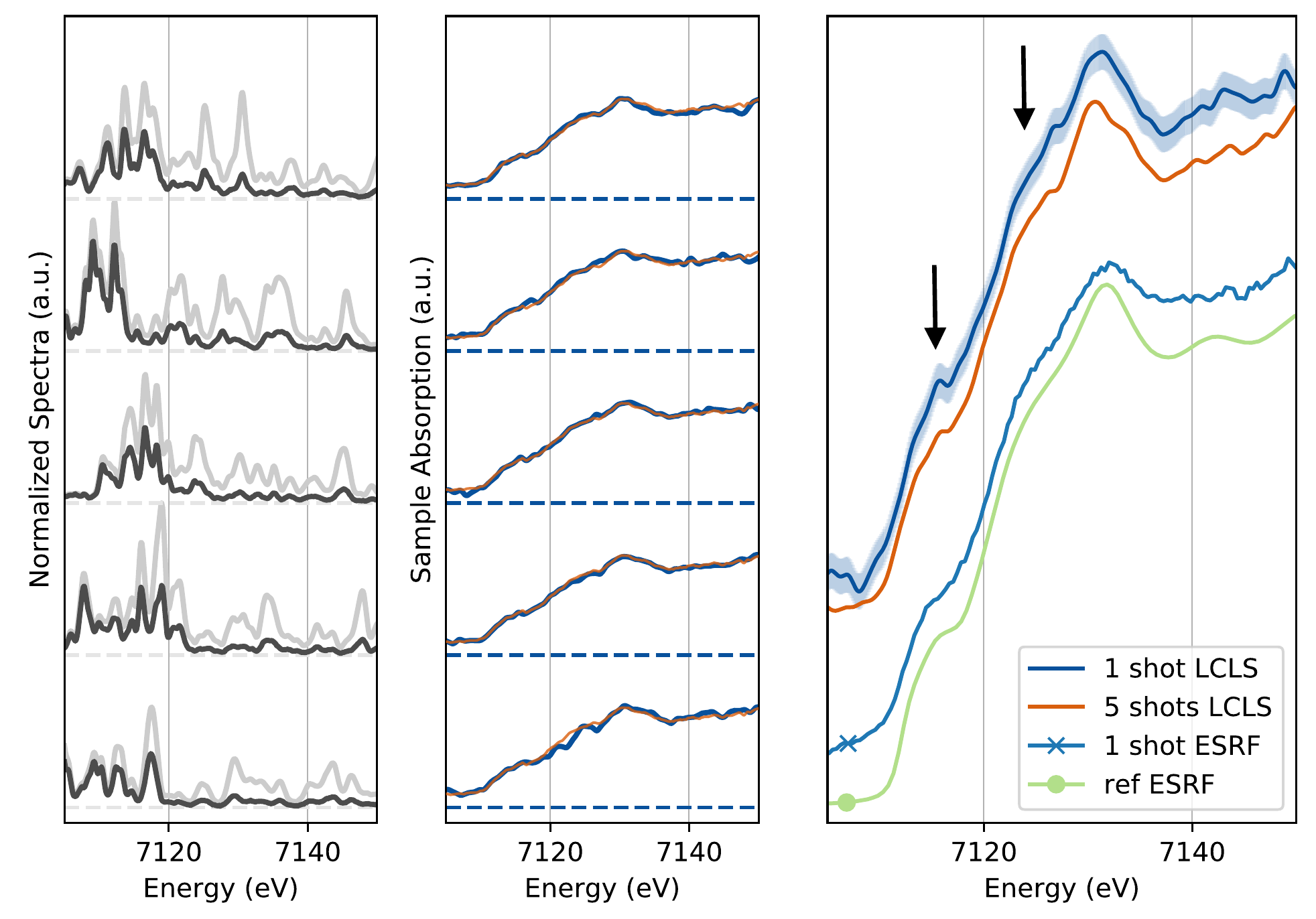}
 \caption{Left panels - XFEL spectra measured before (light gray) and after (black) the sample with the optimized FEL mode for 5 consecutive shots; Middle panels - Corresponding single-shot XANES spectra (blue line) and 5 shots averaged spectra (orange line); Right panel - comparison of different spectra: a single-shot LCLS spectra (blue), 5 shots LCLS averaged (orange), a single bunch ESRF spectra (blue with a cross) and a long time accumulation reference spectra from ESRF (light green with a full dot). The spectra are vertically shifted to improve the clarity of the figure. The two arrows are showing the pre-edge and shoulder distinctives features of the iron K-edge XANES.}
 \label{XANES_BestMode}
\end{figure*}

To further assess the quality of the absorption spectra measured with the dispersive setup, we have measured absorption spectra of a 4 $\mu$m thick iron foil (Goodfellow)  and compared with dispersive data from ESRF \cite{Pascarelli2016, Torchio2016, Mathon2015}. For comparison, an undulator peak from a third generation synchrotron has a continuous spectrum with a minimum bandwidth of several hundred eV \cite{Pascarelli2016}. In figure \ref{XANES_BestMode}, we show five consecutive spectra and their associated sample absorption, convolved with 0.3~eV resolution. Here, the Fe K-edge is clearly observable in a single shot as well as the well identified pre-edge and shoulder features at 7.115keV and 7.125keV respectively \cite{DeBeer2008}. Averaging over the five shots in figure \ref{XANES_BestMode} shows clearly an improvement of the signal to noise and highlights the stability of the single shot data. A single-shot and a long time accumulation XANES spectra taken at ESRF - ID24 are also shown for comparison \cite{Torchio2016, Mathon2015}. The good agreement between ESRF and LCLS data confirms that high quality data can be obtained using an XFEL with one or a few shots despite the unstable, jagged structure of the X-ray spectrum produced by the SASE process.

\begin{figure*}[!h]
\centering
\includegraphics[width=5.5in]{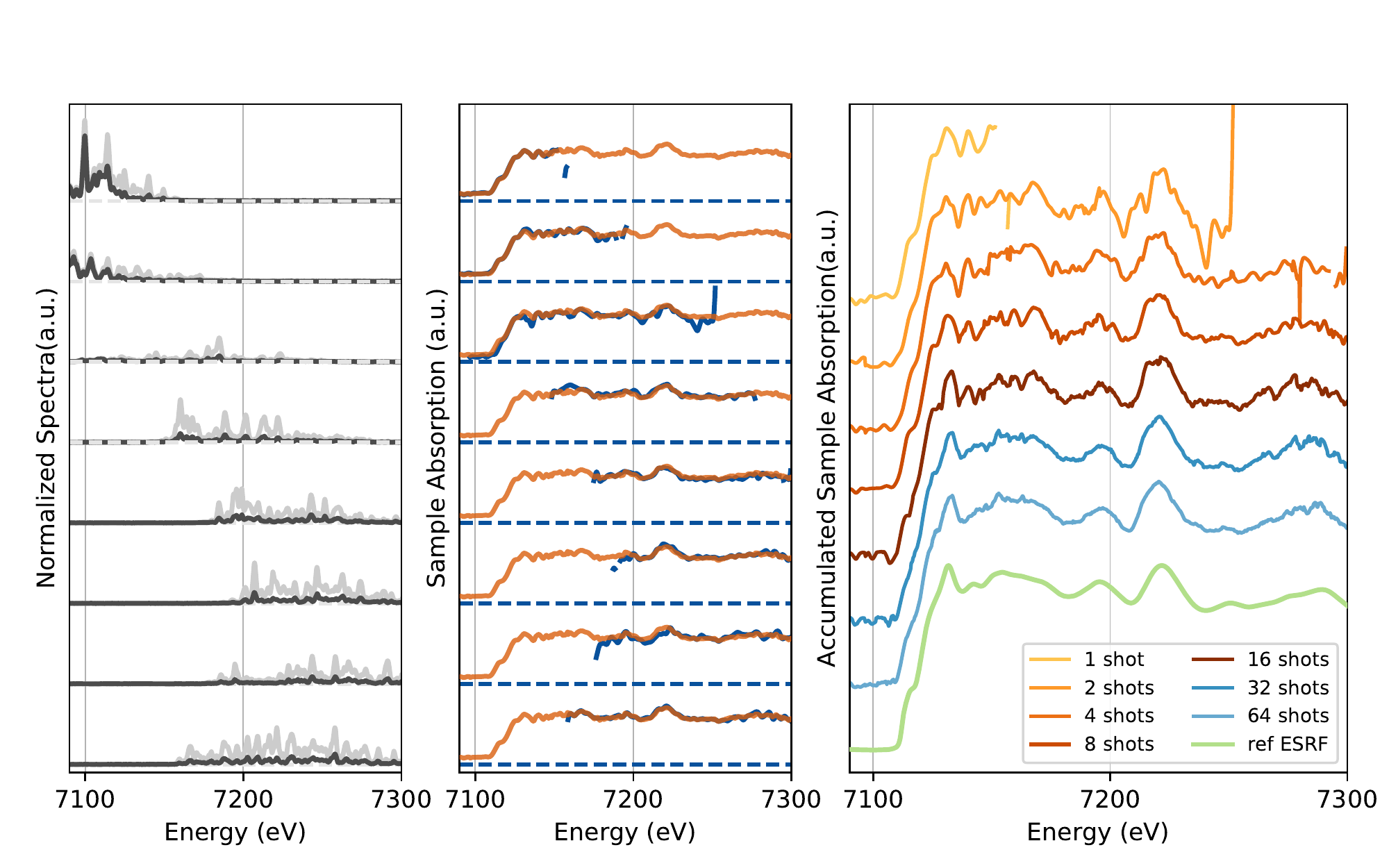}
\caption{Left panels - XFEL spectra of the `best mode' coupled with a controlled photon energy drift to enlarge the average spectral range. The two spectra from the incident and transmitted spectrometers are represented in gray and black respectively. Middle panels - Single-shots (blue line) and 10 shots averaged XANES spectra (orange line). Right panel - Absorption spectra obtained for 1 to 64 accumulated shots compared with a long time accumulation EXAFS spectra from ESRF (light green). }
 \label{EXAFS}
\end{figure*}

Despite the improved single shot spectral window ($\sim$100~eV) obtained with the optimized XFEL settings, an extended spectral range is necessary for certain applications \cite{Ping2013, Filipponi2001,Pettifer2005}. This can be done by changing the central wavelength of the emitted XFEL spectrum. Note also that at LCLS at least, this can be done by the user in a matter of a fraction of a second within 100-200 eV.
To test that our experimental approach was still valid in this case, we have changed the central energy by ~100 eV over the data collection of several shots. Accumulated EXAFS spectra are shown for 1, 2, 4, 8, 16, 32 and 64 shots and are again compared with a typical long-time-accumulation spectra from the ESRF synchrotron. Our results demonstrate that EXAFS features can be reconstructed with only a few shots and that averaging over 10-20 shots allows access to detailed spectral features over an extended energy range. 

\begin{figure}[htb]
\centering
 \includegraphics[width=3in]{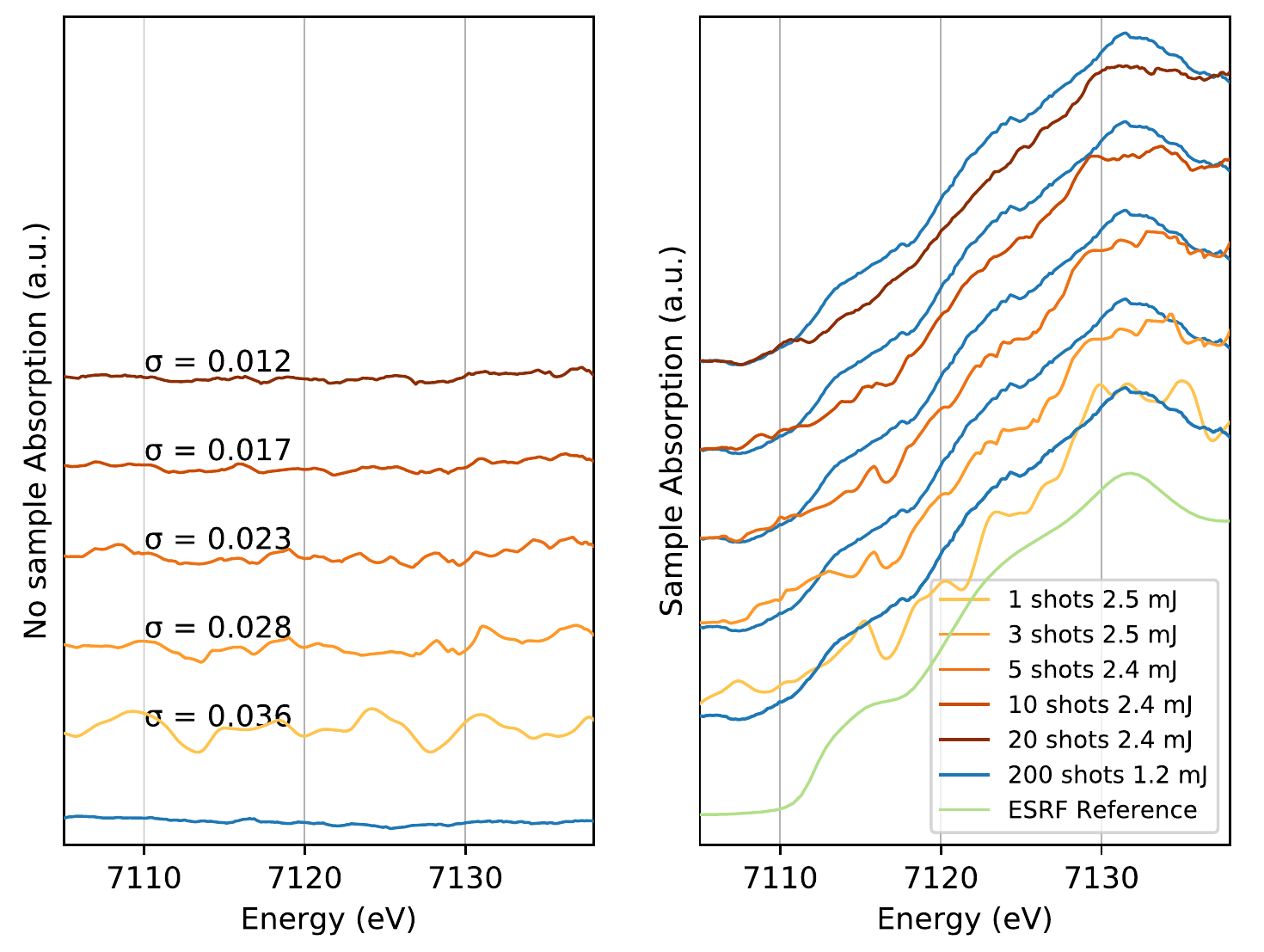}
 \caption{Absorption obtained without a sample (Left) and with a 4$\mu$m Fe sample for a focused XFEL. The XFEL spot size is estimated around 3$\mu$m. The number of accumulated shots is varying from 1 to 20 and the respective standard deviations of the absorption without sample $\sigma$ are varying from 3$\%$ to 1$\%$. On the right panel, the accumulated reference spectra from ESRF is shown in light green and compared with an accumulation of 200 shots at low XFEL flux from the present experiment (blue).}
 \label{Focus}
\end{figure}

Finally, we also evaluated the capability of measuring a single-shot XANES spectra in a focused beam geometry that is usually needed to reach x-ray focal spot smaller than $\sim$~100$\mu$m. In this case, the two spectrometers are symmetrically positioned up and down stream of the beam focus. In this case, in order to preserve the spatial-energy mapping relation in the spectrogram, we needed to flip the scattering direction, as shown in figure \ref{Setup}-(B). We used compound refractive lenses with 4 meters focal length and with the best focus set close to the sample position. Indeed, a smaller beam size at the spectrometer positions implies selecting some part of the spectra at the edge of the XFEL beam itself and being further impacted by spatial aberrations and fluctuations. Despite these difficulties, the obtained XANES spectra are presented in figure \ref{Focus} without and with sample (left and right respectively). This demonstration is particularly relevant as we expect this to be a more common experimental configuration. Indeed, most of the experiments only probe a small fraction of the excited sample as the probed region is usually required to be smaller than the pumped one. For example, the sample can be in the form of a 10 $\mu$m scale diameter liquid jet \cite{Fletcher2016}, or when the uniformly excited region of the sample that is to be probed is well below 100 $\mu$m, limited by the pump laser power and spot size \cite{Nagler2015}. This is also particularly relevant for extreme high pressure and laser heated Diamond Anvil Cell studies that are usually performed with few tens of micron samples \cite{Pascarelli2016,Morard2018}.

In figure \ref{Focus}, we show absorption spectra obtained for 1, 3, 5, 10, 20 and 200 accumulated shots with a few micron focused geometry. The single-shot noise varies between 3 and 5 $\%$ and still achieves around 1$\%$ noise after accumulation. We observe a broadening of the Fe K-edge and a loss of feature of the absorption structure at 7130 keV induced by the XFEL heating. Such disappearance of the Fe pre-edge structure has previoulsy been observed under extreme pressure and temperature conditions at LCLS (unfocused geometry)  \cite{Harmand2015} and at ESRF \cite{Torchio2016}. This change in the XANES spectra is currently used as a diagnostic to detect liquid iron and is attributed to a change of the electronic configuration upon melting. While those mentioned experiments used a secondary driver to melt the iron such as laser compression or laser heating, here, we observe the specific case of self-heating process, meaning that the XFEL probe pulse itself heats the sample. It was also observed by Yoneda et al. \cite{Yoneda2014} at the SACLA facility by accumulating 100 selected shots sorted out by XFEL intensity. By using our experimental set-up in a focusing mode, it only requires the accumulation of few shots to observe this ultrafast process. The quality of our XANES measurement in the focusing geometry could be easily improved by setting up a specific alignment procedure of the spectrometers in this geometry, similar to the one described above for the unfocused beam. This was not performed for this experiment and would have to be repeated for each Compound Refractive Lens (CRL) position. It is important to stress that, when studying ultrafast electronic effects by observing fine features in the XANES spectra, such focused geometry would then require attenuation of the XFEL photon beam intensity to prevent self-heating and ensure a non damaging probe.

\section{Discussion}

While it might have been assumed that averaging data to obtain a smooth incident spectra is an absolute requirement to provide X-ray absorption spectroscopy at XFELs, the presented results show that with the present spectrometer design, refined alignment procedure, and optimized machine parameters, X-ray absorption spectra can be measured with a single FEL pulse. The combination of over compression of the electron bunches with additional tuning of the 'spatial chirp' of the spectra provided single-shot XANES spectra with a few percent noise. In that operation mode, the XFEL bandwidth corresponds to almost 2$\%$. It is important to note that this operation mode also implies a longer pulse duration of the XFEL, on the order of around 50fs. The spectral resolution is given by the thin-bend crystal spectrometers, $\sim$ 0.5eV in the chosen Si(220) configuration. Higher resolution is possible by adopting a different crystal reflection \cite{Zhu2012}. The tuning of the XFEL mode has been performed directly on the X-ray absorption spectra in order to minimize the fluctuations of the absorption, allowing a few percent noise in a single-shot across the primary energy range. Our experimental design sets the path for EXAFS and future finer XANES measurements such as pre-edge modifications of the Fe K-edge following spin transitions \cite{Sanson2016}, valence state quantification \cite{Wilke2001} and reduction of iron in exotic high pressures phases \cite{Boulard2019}, usually requiring signal variations between few $\%$ and down to less than 0.5$\%$. Such typical features of the Fe K-edge would be now clearly visible on a single-shot, or almost, enabling further studies under irreversible conditions at XFELs. However, we should mention this method is only possible when using a transmission geometry and is not suited for fluorescence yield measurements, usually favored for dilute solutions or extremely thin samples.

Our results are particularly important for extreme conditions applications, for which the major role of electrons on atomic structures interplay as been largely identified \cite{Pickard2010}.  Because electronic structure transformation is active on the femtosecond timescale, ultrafast time-resolved X- ray absorption spectroscopy holds great potential for further understand material properties at extreme conditions \cite{Dorchies2016}. This is especially true when studying iron oxides \cite{Wilke2001} and melting curves under high-pressure \cite{Morard2018, Boccato2017}, anomalous heating process such as bond hardening or bond weakening \cite{Gaudin2014,Recoules2006}, electron~-~ion relaxation \cite{Cho2016, Mahieu2018} and ionisation potential depression in warm dense matter \cite{Vinko2012}. Shock-compressed matter or ultrafast laser irradiation are systematically destructive experiments and their reproducibility are extremely sensitive to laser parameters and stability, sample qualities, or even alignment procedure. To date, the approach to managing this issue has been to average XAS spectra taken at similar conditions by ensuring parallel in-situ measurements for post facto sorting. This was performed at the cost of important error bars and loss of data and the presented approach puts an end to this critical experimental  limitation \cite{Harmand2015}. Furthemore, the application of high intensity magnetic field also seeks for high-quality single-shot XANES and X-ray Magnetic Circular Dichro\"{i}sm (XMCD) measurements. The fs pulse duration of the XFEL ensures as well the possibility to track unprecedented transient dynamics. Our results open promising perspectives on the possibility to perform fs time-resolved fine X-ray absorption spectroscopy on highly transitory and challenging systems in order to study the interplay between the dynamic of electronic and atomic structures at extreme regimes of matter. \\

\section { Methods}
{\bf Details on the experimental setup}
Each spectrometer is composed of a 10 $\mu$m thick (around 62$\%$ transmission at 7.1 keV) (220) Si membrane crystal as an analyser and a YAG screen coupled with an optical microscope and a CCD camera to record the spectra on a shot-to-shot basis \cite{Zhu2012}. At 7.1keV, the spectral resolution is $\sim$0.5eV, well below the core hole lifetime broadening of 1.25 eV for the iron K-edge \cite{Krause1979}.
The focusing geometry consists in using CRL lenses and in flipping the second spectrometer in order to keep the 2 systems identical, i.e. the second spectrometer must select the same part of the beam than the first one which is reversed due to focusing beam. \\

{\bf Alignment of Spectrometers}
A fine alignment is performed by online minimisation of the no-sample absorption spectra fluctuations when scanning the first spectrometer position and angle to match the second spectrometer alignment. Figure SM1 in the supplementary material shows the integrated intensity of the second spectrometer while scanning the Bragg angle of the first spectrometer.  We notice a sharp extinction of the signal that corresponds to the expected wavelength extinction at best alignment. To avoid complete extinction of the signal and poor quality measurements, we choose an optimized Bragg angle slightly away from this extinction, indicated by an arrow in the supplementary material. \\

{\bf Analysis} 
The python analysis code is freely available at \href{https://github.com/marcocamma/dispersiveXanes}{https://github.com/marcocamma/dispersiveXanes}. It includes tools to read and analyze the data as well as the scripts used to produce the figure of this manuscripts.. Figures SM2 and SM3 in the supplementary materials shows 2D images of the two spectrometers and their corresponding integrated spectra. We also show the difference of the 2D images and the ratio of the spectra with their calculated standard deviation in the considered spectral range. Figure SM2 shows the results for a single-shot before image transformation while figure SM3 shows the same single shot data but after data treatment. In the figures, we also overlay both spectra (red and blue respectively) to underline their differences before and after image transformation. Once the image transformation is applied to the data, we can observe that the 2 spectra are very well matched. This analysis results in a significant improvement of the absorption fluctuations. For this specific shot, the standard variation $\sigma$, with no sample inserted in the beam, varies from 0.71 to 0.04 after data treatment. \\

{\bf Author Contribution}
MH, AK, HTL, DZ, MCa designed and build the experimental setup and run the experiment with the help of MCh. HTL, DZ, MCa developed the on-line analysis. MH and MCa analyzed the data and wrote the manuscript with help from all other authors\\

{\bf Aknowldgement}
The presented results were obtained at the XPP beamline of the LCLS facility at SLAC. In particular, we would like to thanks the support of the machine experts who optimized the XFEL beam spectral parameters for X-ray absorption spectroscopy.
This work was supported by the French Agence Nationale de la Recherche with the ANR IRONFEL 12-PDOC-0011. This project has also received funding from the European Research Council (ERC) under the European Union Horizon 2020 research and innovation program (grant agreement No. 670787 D PLANETDIVE). AK is currently working under the auspices of the U.S. Department of Energy by Lawrence Livermore National Laboratory under Contract DE-AC52-07NA27344.
We also would like to thank Y. Ding and O. Mathon for helpful discussions. \\

\end{document}